# THE REGENERATIVE VIBRATIONS INFLUENCE ON THE MECHANICAL ACTIONS IN TURNING

Claudiu-Florinel BISU, Raynald LAHEURTE, Alain GERARD, Jean-Yves K' NEVEZ

**Abstract**: In manufacturing during the cutting process the appearance of vibrations can not be avoided. These vibrations constitute a major obstacle to obtain a greater productivity and a better quality of the workpiece. It is thus necessary to develop models which make it possible to study, the totality of the 3D dynamic phenomena. Thanks to an experimental approach the complete torque of the mechanical actions is measured in the presence of the vibrations during a turning operation. This study makes it possible to analyze the variations of the central axis which is a significant parameter of the vibratory phenomena.

## 1. INTRODUCTION

The concepts of the mechanical action (forces and moments) are directly related to the work, led to the beginning of the century past, on the mathematical tool "*torque*" [1]. The need for knowing and for controlling the cutting phenomena is realized by the *torque* analysis. Moreover, the measurement of the mechanical torque is possible thanks to the recent means of metrology [2].

The dynamic phenomena of the machine tools are generated by the interaction of the elastic system machine/cutting process. This interaction represents the dynamic system of the machine tool. The cutting actions implemented to the elastic system cause the relative displacements tool/part, which generate vibrations. Those vibrations influence the section of the chip, the contact pressure, the speed of relative movement etc. Consequently, the instability of the cutting process can cause the instability of the dynamic system of the machine tool. Vibrations appear and have reflected undesirable on the quality of the workpiece surfaces and the wear of the tool. They can generate problems of maintenance even ruptures of elements of machine tool.

Also, it is necessary to develop models making it possible to study the vibratory phenomena met during machining and to provide for the stability cutting conditions the dynamometer with 6 components [2] makes it possible to measure the whole of the mechanical actions forwarded by the mechanical connection between the machined matter (chip and workpiece) and the cutting tool. Measurements then reveal the presence of moments to the point of the tool, not evaluated by the conventional measuring equipment.

The originality of this work relates to the analysis of the moments due to the cut and the influence of the vibrations on the torque implemented to the point of the tool, with the purpose of making evolve a semi analytical 3D cutting model [3].

## 2. EXPERIMENTAL SETUP

The first information which we seek to measure in the formation of the chip is the mechanical actions generated by the workpiece on the tool.

The use of a 6 components dynamometer makes it possible to measure the forces and the moments at the tip of the tool (Fig. 1) and thus to know the complete torque (Fig. 2). The experimental protocol is designed to measure accelerations according to the three cutting directions, thanks to an accelerometer 3D. The area of regenerative vibrations studied is thus identified around 195 Hz [4].

## 3. ANALYZE MECHANICAL ACTIONS

For each test, the complete torque of the mechanical actions are measured then transported with the point of the tool using the conventional relations of transport of the moments.

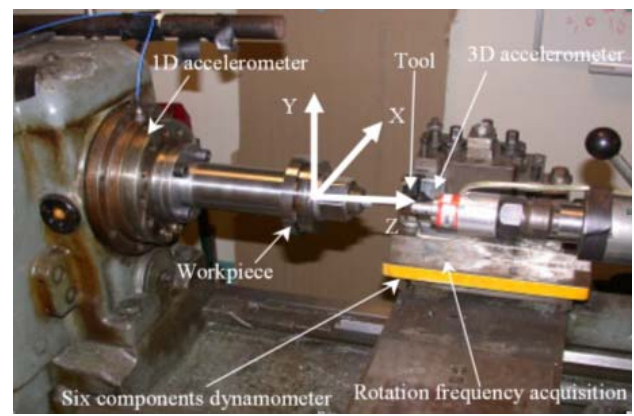

Fig.1. The experimental setup of cutting vibrations.

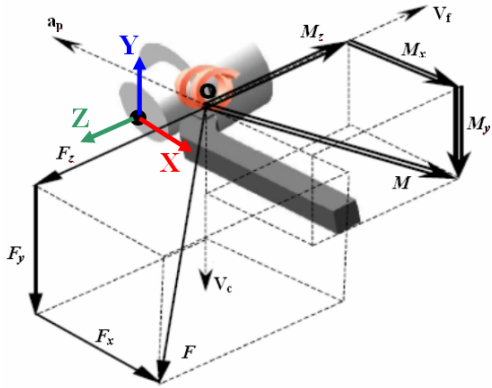

Fig. 2. Definition of the six components; references.

Two series of tests are realized. The first series is carried out with variations of the depth of cut (*ap*) but by maintaining the same value in feed rate (*f*) and the same rotational speed of the workpiece (*N*); the second series takes place with imposed values of *ap*, *N* and variation of the feed rate *f*. For these tests, the tool used is of type TNMA 160412 (nuance carburizes AC700G). The machined material is 42CrMo4. The test-workpiece, of cylindrical form, has a diameter of 120 mm and a length of 30 mm. Then, the regenerative vibrations effect on the mechanicals actions (forces, moments).

### 3.1 Force vector

Using our measuring equipment of the mechanical actions one obtains the signals of the components of the resultant of the forces (Fig. 3). The measured forces make it possible to highlight a development of the variable cutting force Fv around a nominal value Fn [4]. This variable force is a revolving force (Fig. 4) which generates displacements (u) of the tool and the vibrations of the elastic system.

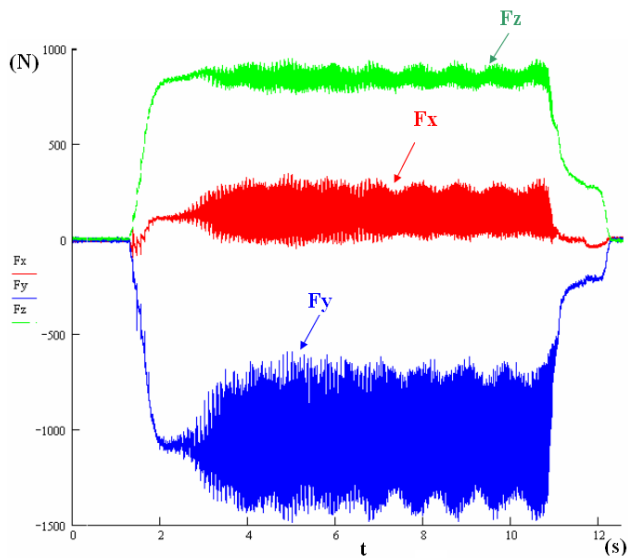

Fig. 3. Signals of the components of the resultant on the three *cutting* directions in the case: ap = 5mm, F = 0.0625 mm/tr and N = 690 tr/min.

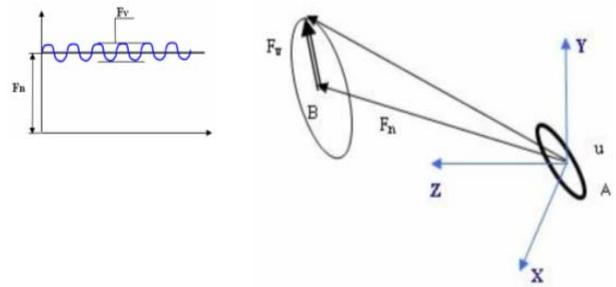

Fig. 4. Development of the cutting force Fv around its nominal value Fn.

This variable force appears in the presence of vibrations and causes the variation of section of the chip [5]. Thus, the variable cutting force and the regenerative vibrations in the elastic system are interactive. The variable force is characterized in two different situations.

These two situations are: the regime of stable cut with a depth cut of 2 mm (without vibrations – Fig. 5a-) and the regime of unstable cut (vibrations) with a depth of cut of 5 mm; we observe the effects of the vibrations on the evolution of the variable forces (Fig. 5b).

The analysis of the tests shows that in the vibratory regime the forces describe an ellipse in the plan (X, Y). This plan is perpendicular to the axis of the cylinder which is not the case in the stable regime (without vibrations). The existence of this plan will allow us to express the dynamic model in the reference marks associated with this plan [4].

### 3.2 Moment vector

The second part of this study is devoted to experimental modeling of the moments in the presence of the regenerative vibrations. During cutting process, torques measurements lead to the acquisition of the moments at the tip of the tool (Fig. 6).

From the signals collected, by mathematical tools the behavior of the moments in vibratory regime is analyzed, while transporting those from the tip of the tool to the central axis.

Indeed, with any torque it is possible to associate a central axis (except the pure couples torques), exclusively calculated starting from the six components of the torque.

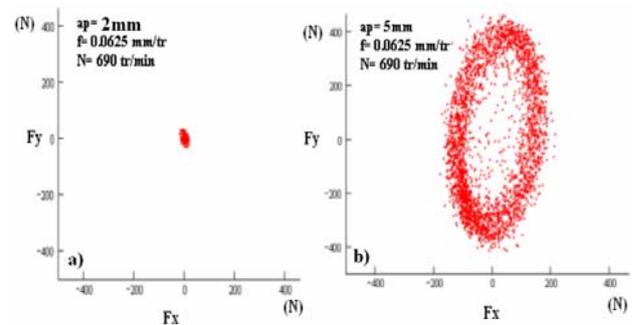

Fig. 5. Stable regime (a); unstable regime (b).

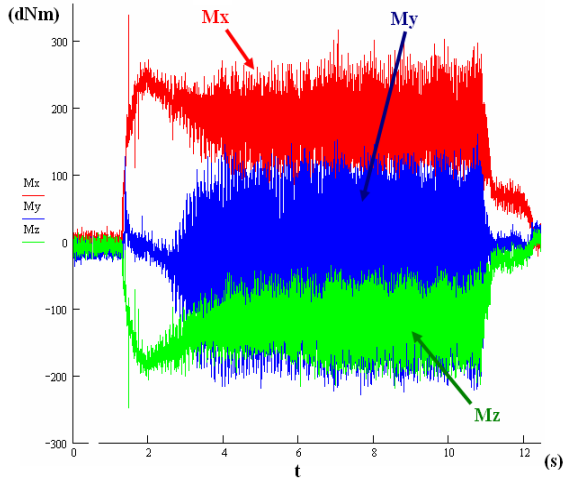

Fig. 6. Signals of the components of the moment on the three cutting directions in the case: ap = 5mm, f = 0.0625 mm/tr and N = 690 tr/min.

The torque in a point O is composed of a resultant force $R$ one moment $M_O$ (1):

$$[A]_O = \begin{cases} \vec{R} \\ \vec{M_O} \end{cases} \quad (1)$$

The central axis is a line defined by a point of reduction and one direction (2):

$$\vec{OA} = \frac{\vec{R} \wedge \vec{M_O}}{\|\vec{R}\|^2} + \lambda \vec{R}, \quad (2)$$

O is the point where the torque of the mechanical actions was moved (tip of the tool) and A is the current point describing the central axis. OA is the vector associated with the bipoint [O, A].
This line (Fig. 7a) corresponds at the points where the moment of the torque of the mechanical actions is minimum. The calculation of the central axis thus amounts determining the whole of the points (a line) where the torque can be expressed according to a force (direction of the straight line) and a pure couple [6].
The central axis represents also the place of the points where the resultant is collinear at the minimum moment. The examination of the six components of the torque of the mechanical actions shows that the average of the forces and the moments are not null.

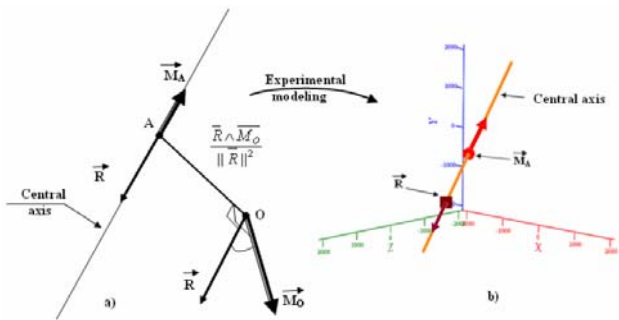

Fig. 7. Representation of the central axis (a); colinearity on the central axis between minimal resultant and moments (b).

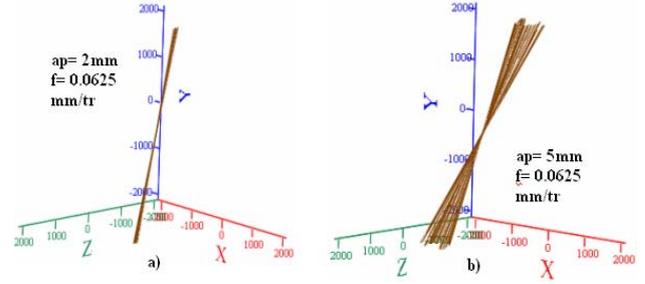

Fig. 8. Representation of the central axes on workpiece rotation in the stable regime (a) and the vibratory regime (b).

For each point of measurement, the central axis is calculated, in the stable regime (Fig. 8a) and unstable (Fig. 8b).
In the presence of vibrations, one can observe the dispersive character of the beam of central axes compared to the stable regime, where this same beam is more tightened, and less tilted compared to the normal one in the plan (X, Z). This dispersion of the central axis can be allotted to the regenerative vibrations which cause the generation of variable moments. While transporting the moment of the point of the tool to the central axis minimum moment $M_A$ is obtained:

$$\vec{M_A} = \vec{M_O} + \vec{AO} \wedge \vec{R}. \quad (3)$$

The test results make it possible to check for each point of measurement the colinearity between the resultant force and the moment calculated at the central axis (Fig.7b).
To express the results of the study, we consider the contact tool/workpiece/chip according to three orthogonal plans (Fig. 9); the tangent plan with the generator of the cylinder ($T_{zy}$) corresponding to the plan (Z, Y), the vertical normal plane ($NV_{xy}$) to represent by (X, Y) orthogonal at the feed cut, and the horizontal normal plane ($NH_{xz}$) characterized by (X, Z) orthogonal at the cutting speed.
From these values of the moment to the central axis we deduce the constant part and the variable part. As for the forces, the variable part is due to the regenerative vibrations as showed it below.
Using this decomposition one can express the attribution of the moments on the areas of contact tool/workpiece/chip. The observations resulting from the analysis show that the vibrations generate rotations of the tool, cause the variations of contact and thus generate variable moments. This representation enables us to express the moments following the three plans: moment of orthogonal swivelling in the tangent plan ($T_{yz}$) and two moments of bearing contained in the normal plans vertical ($NV_{xy}$) and horizontal ($NH_{xz}$).
The components of these moments to the central axis show a development quasi-constant of the moment of pivoting according to the variable advances, the moment following X which is perpendicular to the tangent plan (T), while with the point of the tool this component of the variable moment is most important.

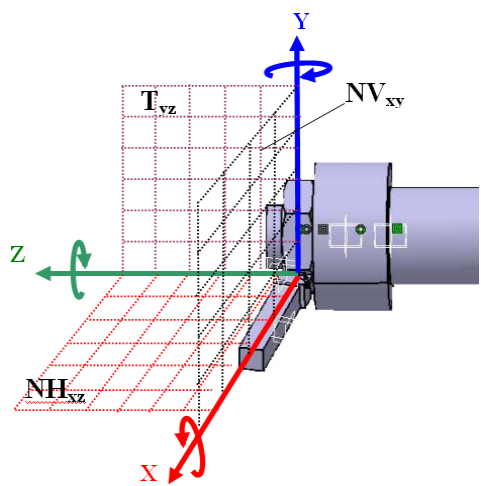

Fig.9. Plans corresponding at the contact tool/workpiece/chip.

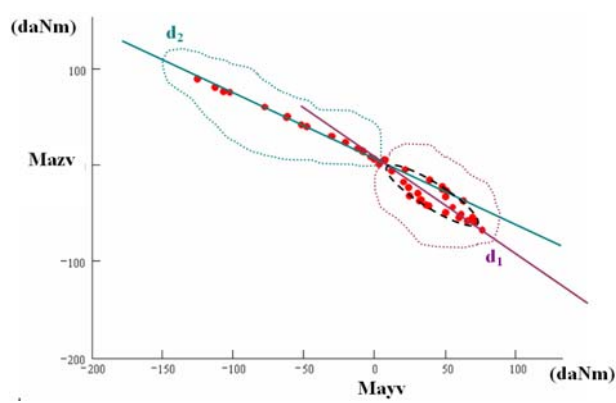

Fig.10. Representations of the moments to the central axis divided into two parties on a rotation of workpiece for ap 5mm, f = 0.0625 mm/tr and N = 690 tr/min.

After the analysis of the components of the moments determined with the central axis one could observe a localization of the moments on two distinct areas. Taking account of this remark, the components of the moments following the three directions are separated into two parts noted d1 and d2 (Fig.10).

In the vibratory case ap = 5 mm, the first family of points d2 is organized according to a line while the second family is area of an pinched elliptic type around a line d1 (the large axis). In the case without vibrations ap = 2mm, we note that the two d1 families and d2 are confused in only one family organized according to only one line. Thus, the appearance of the elliptic family around the line d1 is thus quite due to the regenerative vibrations of the case ap = 5mm. In addition the frequency of the vibrations of the d1 party is higher than that of the d2 party. Moreover, with the central axis, the d1 families and d2 seem to correspond to elements distinct from the generated surface.

## 4. CONCLUSIONS

One could set up an experimental operative paragraph authorizing to analyze the relations resulting vibrations – resultants - moments. This study allows, during a turning operation, establishing correlations between the regenerative vibrations and the central axis of the torque of mechanical actions. It is thus possible, thanks to the use of the parameters defining the central axis to study the development of the vibrating system tools – workpiece. This is a new tool usable to supervise the vibrations at the time of the cut.

The description of the moments at the time of the cut and the pure couples is carried out. A first conclusion on the influence of the vibrations concerning the moments is allotted to the variations of contact tool/workpiece/chip. Measurements on the dynamic behavior of the tool during the cut thanks to two accelerometers 3D must enable us to establish the position of the tool according to time and to look further into this first conclusion.

Another prospect is to supplement the measurement of displacements by the measurement of the swing angles of the tool with the purpose of building the matrix of complete rigidity of the tool and the part.

Thereafter, having the complete torque, a model will be proposed taking into account the overall behavior of surface rubbing with the purpose to identify the influence of the vibrations on the tribological behavior of the tool/chip contact.

**REFERENCES**

[1]. Stawell R.Ball (1900). *A treatiste on the theory of screws*, At the university press, Cambridge.
[2] Couétard Y., *(2000).Caractérisation et étalonnage des dynamomètres à six composantes pour torseur associé à un système de forces*, Thèse, Université Bordeaux 1.
[3] Cahuc O, Darnis O., Gérard A., Bataglia JL., (2001). *Experimental and analytical balance sheet in turning application,* The international Journal of Advanced Manufacturing Technology, Vol. 18, N°9, pp.648-656.
[4] Bisu C., Darnis P., K'nevez J.Y., Cahuc O., Laheurte R., Ispas C., (2006). *Un nouveau modèle expérimental d'analyse des phénomènes vibratoire lors d'une opération de tournage*, 4èmes Assises MUGV, Aix en Provence, 8-9 juin 2006, France.
[5] Marinescu I., Ispas C., Boboc D., (2002). *Handbook of Machine Tool Analysis*, Marcel Deckker, Inc, ISBN 0-8247-0704-4, USA.
[6] Laporte S.,(2005), *Comportement et endommagement de l'outil en perçage à sec: applications aux assemblages aéronautiques,* Thèse, Université Bordeaux.

**Authors:**

Claudiu-Florinel BISU, Doctorant, Université Bordeaux1, Laboratoire de Mécanique Physique UMR CNRS 5469, et LMSP, Université Polytechnique de Bucarest, Splaiul Independentei 313 Bucarest – Roumanie, E-mail : cbisu@u-bordeaux1.fr
Raynald LAHEURTE, Maître de conférence, IUT Bordeaux1, Laboratoire de Génie Mécanique et Matériaux de Bordeaux, et Laboratoire de Mécanique Physique UMR CNRS 5469 Université Bordeaux1 E-mail : raynald.laheurte@u-bordeaux1.fr
Alain GERARD, professeur des universités, Université Bordeaux 1, Laboratoire de Mécanique Physique UMR CNRS 5469, E-mail : alain.gerard@u-bordeaux1.fr
Jean-Yves K'NEVEZ, Maître de conférence, Université Bordeaux 1, Laboratoire de Mécanique Physique UMR CNRS 5469, E-mail: jean-yves.knevez@u-bordeaux1.fr